\documentclass[pdflatex,sn-mathphys-ay]{sn-jnl}

\usepackage{graphicx}%
\usepackage{multirow}%
\usepackage{amsmath,amssymb,amsfonts}%
\usepackage{amsthm}%
\usepackage{mathrsfs}%
\usepackage[title]{appendix}%
\usepackage{xcolor}%
\usepackage{textcomp}%
\usepackage{manyfoot}%
\usepackage{booktabs}%
\usepackage{algorithm}%
\usepackage{algorithmicx}%
\usepackage{algpseudocode}%
\usepackage{listings}%
\usepackage[utf8]{inputenc}

\theoremstyle{thmstyleone}%
\theoremstyle{thmstyletwo}%

\theoremstyle{thmstylethree}%
\newcommand{\argmax}{\mathop{\rm arg~max}\limits}

\begin{document}

\title{Incorporating the underuse problem in the tragedy of the commons}


\author*[1]{\fnm{Shota} \sur{Shibasaki}}\email{sshibasa@mail.doshisha.ac.jp}

\author[2]{\fnm{Wakaba} \sur{Tateishi}}

\author[3]{\fnm{Shuhei} \sur{Fujii}}
\author*[4, 5]{\fnm{Ryosuke}\sur{Nakadai}}\email{nakadai-ryosuke-pt@ynu.ac.jp}

\affil[1]{\orgdiv{Faculty of Culture and Information Science}, \orgname{Doshisha University}, \orgaddress{\street{1-3 Tatara Miyakodani}, \city{Kyotanabe}, \postcode{610-0394}, \state{Kyoto}, \country{Japan}}}

\affil[2]{\orgdiv{Department of Business Administration}, \orgname{Hokkaido Musashi Women’s University}, \orgaddress{\street{Kita 22-Jo Nishi 13-Chome, Kita-ward}, \city{Sapporo}, \postcode{001-0022}, \state{Hokkaido}, \country{Japan}}}

\affil[3]{\orgdiv{The Institute for Japanese Culture and Classics}, \orgname{Kokugakuin University}, \orgaddress{\street{4-10-28 Higashi}, \city{Shibuya-ward}, \postcode{150-8440}, \state{Tokyo}, \country{Japan}}}

\affil[4]{\orgdiv{Faculty of Environment and Information Sciences}, \orgname{Yokohama National University}, \orgaddress{\street{79-7 Tokiwadai, Hodogaya-ward}, \city{Yokohama}, \postcode{240-8501}, \state{Kanagawa}, \country{Japan}}}

\affil[5]{\orgname{Research Institute for Humanity and Nature}, \orgaddress{\street{457-4 Motoyama, Kamigamo}, \city{Kyoto}, \postcode{603-8047}, \state{Kyoto}, \country{Japan}}}


\abstract{


The tragedy of the commons has traditionally been framed as a problem of
resource overuse driven by self-interested exploitation. In contrast, growing empirical evidence shows that insufficient use or abandonment of natural
resources, known as underuse, can also lead to ecological degradation and loss
of ecosystem services. Despite its relevance, underuse has rarely been examined
within evolutionary theories of resource use. Here, we develop a simple eco-evolutionary model that integrates both provisioning and non-provisioning ecosystem services to analyze the evolution of resource-use strategies. Using adaptive dynamics, we investigate how individual resource use evolves while altering resource abundance. The model shows that overuse and underuse arise naturally as alternative evolutionary outcomes of the same underlying process, alongside intermediate use and evolutionary branching.
We derive analytical conditions for the existence, number, and stability of
evolutionarily singular strategies, and show that the qualitative evolutionary
fate is primarily determined by the shape of provisioning benefits. Only when
provisioning benefits increase in a concave manner does evolutionary dynamics
converge to a unique intermediate strategy that is continuously stable. In contrast,
convex increasing benefits generate a broader range of outcomes:
overuse, underuse, bi-stability, and evolutionary branching. By explicitly
comparing the continuously stable strategy with the socially optimal strategy,
we further quantify how their deviations depend on the valuation of
non-provisioning services.
Our results provide a theoretical framework for viewing the common-pool resource
dilemmas as intrinsically two-sided evolutionary problems, and offer a baseline
for future studies exploring interventions to address overuse and underuse simultaneously.
}


\pacs[Statements and Declarations]{
The authors declare no competing interests.}
\pacs[ORCID]{
\\
S. S. 0000-0002-8196-0745. \\
W. T. 0009-0004-3589-8137.
\\
S. F. 0000-0002-4777-7752.
\\
R. N. 0000-0002-9512-8511.
}
\maketitle

\section{Introduction}
Human societies have long depended on a wide variety of natural resources, including forests, fisheries, and grassland. These resources are typically shared and renewable, functioning as common-pool resources whose benefits are non-excludable yet whose extraction is inherently rivalrous \citep{Malezieux2025}. This fundamental tension has raised persistent concerns that these resources will be overused by rational individuals, notably articulated in Hardin's ``tragedy of the commons'' \citep{Hardin1968}. Since then, scholars across biology, economics, and psychology have sought to understand how overuse can be avoided, revealing institutional, behavioral, and evolutionary processes that can enable the sustainable use of shared natural resources \citep{Ostrom1990Book, Ostrom2008, Boyd2018Science, Frischmann2019}.
\par
While previous studies have focused predominantly on overuse, recent studies have highlighted that underuse of natural resources has been largely overlooked in the literature \citep{Shimada2018}. Underuse refers to reduced resource-use intensity, or abandonment of resource management or use. A growing body of research shows that underuse can lead to broad ecological degradation. For example, a recent review shows that the abandonment of formerly managed forests reduces biodiversity, alters forest structure, and diminishes multiple ecosystem services \citep{Oono2020}.  Empirical studies documenting and investigating these problems have been particularly prominent in Japan \citep{Kobayashi1998JJSR, Shimada2015, Hirahata2020} and European countries \citep{Baur2018, Baur2021, Brosette2022}. 
\par
\citet{Morino2014} associates the underuse problem with a shift in the expected benefits of ecosystem services \citep{MA2005}: from provisioning  (e.g., obtaining food and fiber) to non-provisioning services (i.e., regulating, supporting, and cultural). A key distinction lies in the nature of the services derived from natural resources. Provisioning services are typically individual benefits derived from extracting units of natural resources. In other words, conventional common-pool resource problems have focused on provisioning services. In contrast, non-provisioning services are non-extractive and shared among members of a society: the benefits arise from the continued existence and ecological condition of the natural resources. Examples of such non-provisioning services include water purification, photosynthesis, and aesthetic values \citep{MA2005}.
Consequently, focusing solely on provisioning services naturally highlights the risk of overuse, whereas incorporating non‑provisioning services reveals that reduced use or abandonment can also lead to substantial ecological losses. This distinction implies that both overuse and underuse are inherent risks in shared renewable resource systems, depending on which ecosystem services are emphasized. Therefore, a theoretical framework that integrates benefits derived from both provisioning and non-provisioning services is required to understand how resource-use strategies evolve under ecological feedback and how overuse and underuse problems can be avoided simultaneously.
\par
Here, we provide a simple mathematical model that incorporates both overuse and underuse problems by explicitly integrating the benefits of provisioning and non-provisioning services. We analyze how the evolution of resource-use strategies alters the amounts of natural resources within the framework of adaptive dynamics \citep{Geritz1997, Geritz1998, Broom2013, Avila2023}. This approach is particularly suited for analyzing the gradual evolution of continuous traits under ecological feedback \citep{Ferriere2012}. In our model, the intensity of resource use directly alters the abundance of the natural resource, and this abundance, in turn, shapes the fitness landscape of alternative exploitation strategies. Thus, our framework explicitly incorporates eco-evolutionary feedback in a manner analogous to recent studies in the evolutionary game theory that couple strategic behaviors with environmental dynamics \citep{Weitz2016, Tilman2019, Shibasaki2020,Shibasaki2025Supernatural}. This feature distinguishes our model from continuous public-goods games applying adaptive dynamics \citep{Johnson2021RSOS, LuoJ2025}, which do not explicitly incorporate resource dynamics.
Within the present eco-evolutionary framework, the discrepancy between the socially optimal resource level and the evolutionarily realized one provides a natural and quantitative measure of the severity of overuse or underuse problems. We derive explicit analytical conditions under which intermediate exploitation levels emerge as continuously stable strategies (CSSs), and we identify parameter regimes in which evolution instead drives populations toward socially undesirable overuse or underuse. We further quantify the extent of these deviations at CSSs and characterize the ecological and evolutionary conditions under which both problems can be mitigated. Our model, therefore, theoretically extends classical ``tragedy of the commons'' problems by incorporating two-sided resource-use problems. Future research is encouraged to use our model as a baseline when addressing how interventions facilitate the evolution of appropriate resource use strategies.
\section{Model and Methods}
\subsection{Formulation}
Suppose a society uses the natural resource $R$, which provides several types of ecosystem services. For the sake of simplicity, we divide the benefits from the natural resource into two categories: the provisioning service $f(u, R)$, which increases over the amount of resource and the intensity of usage ($0 \leq u \leq 1$), and the non-provisioning services $g(R)$, which depends solely on the amount of natural resource.
We consider the maximization of the payoff an individual receives from the natural resource:
\begin{align}
    \phi(u, R) = f(u, R) + g(R) \label{eq:target}.
\end{align}
Specifically, we assume
\begin{align}
    f(u, R)&= b (au R)^w-cu,\\
    g(R) &= d \exp\{-e(R_{\mathrm{opt}}-R)^2\},
\end{align}
where $b$ is the weight of the provisioning service, $a$ is the maximum individual exploitation rate, $c$ is the cost of utilizing natural resource, $d$ is the weight of the non-provisioning services, $e$ control the decrease of these services when the amount of natural resource deviates from the optimal value $R_{\mathrm{opt}}$. 
 The current formulation of $ f(u, R)$ represents broad patterns of increasing functions of the provisioning service by tuning the exponent, $w$: $w= 1$ corresponds to a linear function, $0 < w<1$ corresponds to a concave function, and $w > 1$ corresponds to a convex function.
\par
We assume that the provisioning services increase with $u$ and $R$ because these services are obtained by extracting resource units.
By contrast, the benefits from the non-provisioning services depend solely on $R$, and we assume that it is maximized at a certain level of natural resources, $R_{\mathrm{opt}}$. Because an empirical quantification of non-provisioning services over the amount of natural resources is unavailable, we formulated $g(R)$ so that it can flexibly represent broad patterns by tuning $R_{\mathrm{opt}}$. Formally, $R_{\mathrm{opt}} \leq 0$ indicates that $g(R)$ is a monotonically decreasing function, while it is a monotonically increasing function when $R_{\mathrm{opt}}$ exceeds the carrying capacity (see below): otherwise, $g(R)$ is a unimodal function.
\par
When all individuals apply the identical strategy $u$, the dynamics of the amount of natural resource $R$ is governed by the logistic growth and exploitation by the society:
    \begin{align}
    \frac{dR}{dt} &= \mu R \left(1-\frac{R}{K}\right) - a u NR,  
\end{align}
where $\mu$ is the intrinsic growth rate of the natural resource, $K$ is the carrying capacity of the natural resource, and $N$ is the number of individuals in the society. This equation gives two equilibria: one is trivial, $R=0$, and the other is
\begin{align}
    R^*(u) = K\left(1-\frac{\alpha u}{\mu}\right),\label{eq:equilibrium_R}
\end{align}
where $\alpha=a N$ represents the maximum social exploitation rate. This equilibrium is positive and stable if and only if $\alpha u<\mu$. In this manuscript, we assume that the intrinsic growth rate exceeds the maximum social exploitation rate ($\mu > \alpha$), so that the equilibrium $R^*(u)$ is positive and stable for all $0\leq u\leq 1$. 
\par
At this equilibrium point, Eq \eqref{eq:target} is maximized at $u_{\mathrm{opt}}$:
\begin{align}
    u_{\mathrm{opt}} \equiv \argmax_{u} \phi(u, R^*(u)).
\end{align}
This strategy is socially optimal in the sense that all individuals maximize their benefits when employing the same strategy. Specifically, this manuscript focuses on the case where $0<u_{\mathrm{opt}}<1$, so that both overuse and underuse can occur. If $u_{\mathrm{opt}}=0, 1$, only the underuse and overuse problems can happen, respectively. We evaluate the deviation of the evolved strategy $u$ from the social optimum  $u_{\mathrm{opt}}$. If $u>u_{\mathrm{opt}}$, this represents the overuse problem, while $u<u_{\mathrm{opt}}$ does underuse one. The deviations quantify the degree of these two problems, respectively. This quantification broadens the fourth definition of overuse/underuse in \citet[Table 1]{Ohsawa2019} by explicitly incorporating both provisioning and non‑provisioning ecosystem services.

\subsection{Adaptive dynamics}
We apply the adaptive dynamics \citep{Geritz1997,Geritz1998,Broom2013, Avila2023} to investigate the evolutionary dynamics of $u$. The adaptive dynamics approach assumes that all individuals except one mutant employ the same strategy $u$ and that the amount of natural resource is at equilibrium, $R^*(u)$. When the mutant utilizes the natural resource at rate $v$, the invasion fitness is written as follows:
\begin{align}
    s(v,u) &= \phi\left(v, R^*(u)\right)-\phi\left(u, R^*(u)\right) \nonumber \\
&=b(v^w-u^w)\{aR^*(u)\}^w -c(v-u). \label{eq:invasion}
\end{align}
This equation shows that invasion fitness equals the difference in benefits from the provisioning service. This is because the benefits from the non-provisioning services $g(R(u))$ depend only on the equilibrium resource level determined by the resident strategy $u$. Table \ref{tab:parameters} lists all parameters used in this model.

\begin{table}[t]
\centering
\caption{Parameters and variables used in the model}
\label{tab:parameters}
\begin{tabular}{lll}
\hline
Symbol & Range or value & Description \\
\hline
$R$ &$R\geq 0$  & Amount of the natural resource \\
$u$ & $0 \le u \le 1$ & Resident intensity of resource use \\
$v$ & $0 \le v \le 1$ & Mutant intensity of resource use\\
$a$ &$a>0$  & Maximum individual exploitation rate\\
$b$ &$b>0$  & Weight of the provisioning service \\
$c$ & $c\geq 0$ & Cost of utilizing the natural resource \\
$w$ & $w>0$ & Exponent determining the nonlinearity of $f(u, R)$ \\
$d$ &$d\geq 0$  & Weight of non-provisioning service \\
$e$ &$e>0$  & Strength of decrease in $g(R)$ away from $R_{\mathrm{opt}}$ \\
$R_{\mathrm{opt}}$ &$0\leq R_{\mathrm{opt}} \leq K$  & Resource level that maximizes $g(R)$ \\
$\mu$ & $\mu>\alpha$ & Intrinsic growth rate of the natural resource \\
$K$ & $K>0$ & Carrying capacity of the natural resource \\
$N$ & $N>0$ & Number of individuals in the society \\
$\alpha$ & $\alpha = aN$ & Maximum social exploitation rate \\
\hline
\end{tabular}
\end{table}
\par
Adaptive dynamics investigate the long-term evolution of the resident strategy $u$ under a weak mutation (i.e., $v$ is close to $u$). An evolutionarily singular strategy $u^*$ is written as follows:
\begin{align}
    \left.\frac{\partial s}{\partial v}\right|_{u=v=u^*} &= b\{aR^*(u^*)\}^w w u^{*w-1}-c=0. \label{eq:gradient}
\end{align}
\par
Four properties of an evolutionarily singular strategy $u^*$ are analyzed through the second partial derivative of the invasion fitness \citep{Geritz1997, Broom2013}. First, an evolutionarily singular strategy $u^*$ is called non-invasible \citep{Broom2013} or evolutionarily stable \citep{Geritz1997, Brannstrom2013} if 
\begin{align}
    \left.\frac{\partial^2 s}{\partial v^2}\right|_{u=v=u^*} &= b\{aR^*(u^*)\}^w w(w-1)u^{*w-2}<0. \label{eq:ES}
\end{align}
In other words, such an evolutionarily singular strategy $u^*$ cannot be invaded by any nearby mutant $v$.
\par
Second, an evolutionarily singular strategy is called convergence stable if 
\begin{align}
    &\left.\frac{\partial^2 s}{\partial v^2}\right|_{u=v=u^*}+
\left.\frac{\partial^2 s}{\partial v \partial u}\right|_{u=v=u^*}  < 0, \nonumber \\
\Leftrightarrow 
 &-\alpha w u^* +(w-1)(\mu- \alpha u^*) < 0. \label{eq:convergent}
\end{align}
In other words, directional selection drives the resident strategy toward $u^*$ when the evolutionarily singular strategy is convergence-stable. An evolutionarily singular strategy that satisfies inequalities \eqref{eq:ES} and \eqref{eq:convergent} is called a continuously stable strategy (CSS), which is an endpoint of the evolutionary dynamics \citep{Dieckmann2004, Brannstrom2013}. On the other hand, an evolutionarily singular strategy that is convergence stable but not evolutionarily stable is called an evolutionary branching point, in which the society splits into two subpopulations composed of distinct strategies.
\par
Third, an evolutionarily singular strategy can invade nearby strategies if and only if 
\begin{align}
    \left.\frac{\partial^2 s}{\partial u^2}\right|_{u=v=u^*} &=
    \frac{b w (a R^*(u^*))^w}{\mu-\alpha u^*}\{(3w-1)\alpha u^{w-1}+\mu(1-w)u^{w-2}\}>0. \label{eq:invade_near}
\end{align}
When the above inequality is satisfied, an evolutionarily singular strategy can spread in other populations when it is initially rare  \citep{Geritz1997}.
\par
Finally, an evolutionarily singular strategy is called a protected dimorphism when two strategies on either side of an evolutionarily singular strategy can invade each other. This property is satisfied if and only if 
\begin{align}
    &\left.\frac{\partial^2 s}{\partial u^2}\right|_{u=v=u^*} + \left.\frac{\partial^2 s}{\partial v^2}\right|_{u=v=u^*}
=\frac{2\alpha b w^2 (aR^*(u^*))^{w} u^{* w-1}}{\mu-\alpha u^*} >0.\label{eq:protected}
\end{align}
Note that the above inequality always holds when an evolutionarily singular strategy is $0 < u^* <1$.
\section{Results}\label{sec:results}

\subsection{Existence and number of evolutionarily singular strategies}\label{sec:number_ess}
We first characterize the conditions under which an interior evolutionarily singular strategy exists (i.e., $0<u^*<1$) in our model. We then derive the number of singular strategies that can arise under these conditions. Note that the absence of interior evolutionarily singular strategies results in either overuse or underuse problems because the left-hand side of Eq \eqref{eq:gradient} never becomes zero and because the social optimum $u_{\mathrm{opt}}$ exists between 0 and 1. If the left-hand side of Eq \eqref{eq:gradient} is positive for all $u^* \in (0,1)$, $u^*$ evolves toward $1$, leading to the overuse, while the underuse happens when the left-hand side of Eq \eqref{eq:gradient} is negative for all $u^* \in (0,1)$.
\par
An evolutionarily singular strategy $u^*$ is a root of Eq \eqref{eq:gradient}:
\begin{align}
 \left.\frac{\partial s}{\partial v}\right|_{v=u=u^*}&=0 \nonumber \\
\Leftrightarrow\left(1-\beta u^*\right)^{w}{u^*}^{w-1} -\gamma&=0.
\end{align}
where
\begin{align}
    \beta&=\frac{\alpha}{\mu},\\
    \gamma&=\frac{c}{ bw(aK)^w}
\end{align}
and thus, $0<\beta<1$ and $\gamma \ge 0$. 
Then, $u^*$ is the intersection of the following two curves.
\begin{align}
\left\{\begin{array}{l}
    y=\left(1-\beta u^*\right)^{w}{u^*}^{w-1}=:h(u^*), \\
    y=\gamma.
    \end{array}
    \right.
\end{align}
\par
Importantly,
\begin{align}
    \lim_{u^* \rightarrow 0_+}h(u^*) &= 
    \left\{\begin{array}{cc}
    \infty & (0<w<1)\\
    0 &(w>1),
    \end{array}\right.\\
    h(1)&=(1-\beta)^w>0,\\
    h'(u^*)&=(1-\beta u^*)^{w-1}u^{*w-2}[(w-1)(1-\beta u^*)-\beta w u^*]. \label{eq:gradient_h}
\end{align}
We wrote
\begin{align}
    u^\dagger = \frac{w-1}{\beta(2w-1)}
\end{align}
for simplicity because $h'(u^\dagger)=0$. Since $0<\beta<1$, $u^\dagger$ exists between 0 and 1 if and only if
\begin{align}\left\{
    \begin{array}{l}
    \beta > \frac{w-1}{2w-1} \\
    w>1
    \end{array}\right.. \label{eq:u_dagger}
\end{align}
Below, we classify the situations according to whether the inequalities \eqref{eq:u_dagger} are satisfied, to see how the number of interior evolutionarily singular strategies changes. 
\par
\begin{figure}[tbh]
    \centering
    \includegraphics[width=0.95\linewidth]{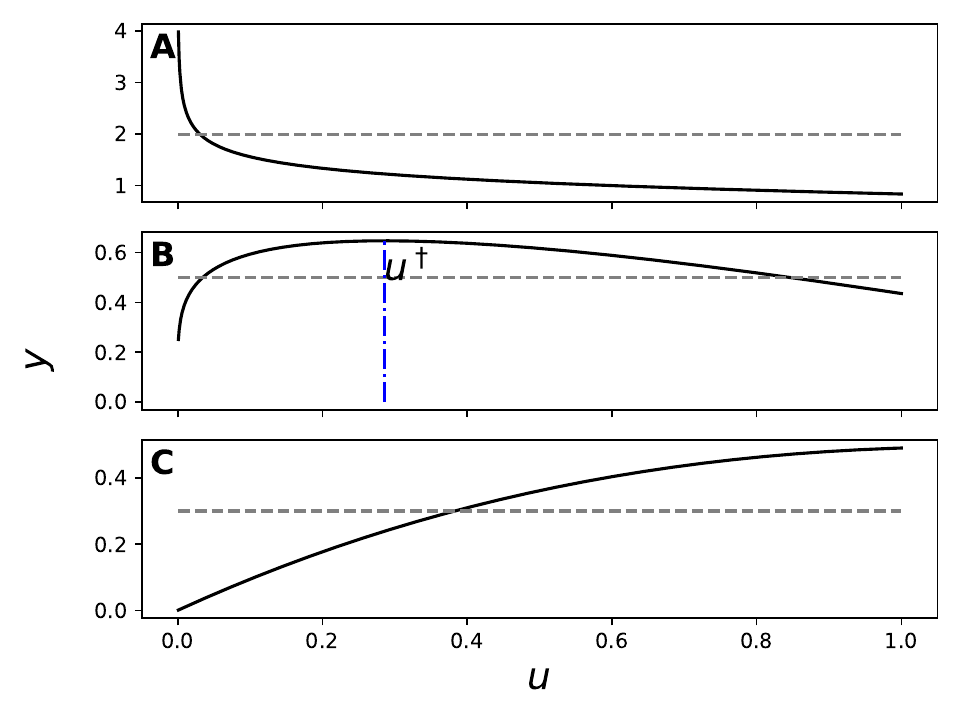}
    \caption{Number of evolutionarily singular strategies}
    \label{fig:number_of_ess}
\begin{flushleft}
    {\small
    The evolutionarily singular strategies $u^*$ are the intersections of $y=h(u^*)$ (the black solid curve) and $y=\gamma$ (the horizontal dashed line). $A$: When $0<w<1$, the number of evolutionarily singular strategies is at most one. B and C: When $w>1$, the number of evolutionarily singular strategies is at most two (if $0<u^\dagger<1$) or one (otherwise). The blue dotted vertical line in panel B represents the $u=u^\dagger$, which gives $h'(u^\dagger)=0$. Parameter values are as follows: $(\beta, \gamma, w) = (0.2, 2, 0.8 )$ in A $(0.5, 0.5, 1.2)$ in B, and $(0.3, 0.3, 2)$ in C, respectively.
    
    }
\end{flushleft}
\end{figure}

When $0<w<1$, $h'(u^*)<0$. Then, a unique $u^*$ exists between $0$ and $1$ (Fig. \ref{fig:number_of_ess}A) if and only if
\begin{align}
     \gamma >h(1)=(1-\beta)^w.\label{eq:ess_exist1}
\end{align}
Otherwise, the left-hand side of \eqref{eq:gradient} is positive for all $u^* \in (0,1)$, resulting in the overuse of natural resources ($u\to 1$).
\par
When $w>1$ and $\beta > (w-1)/(2w-1)$, $h(u^*)$ is a unimodal function over $u^*$; $h'(u^*)>0$ ($0<u^*<u^\dagger$) while $h'(u)<0$ ($u^\dagger<u^*<1$). Thus $u^*$ exists between $0<u^*<1$ if and only if
\begin{align}
 \gamma \leq h(u^\dagger). 
\end{align}
Equality corresponds to $u^*=u^\dagger$.
The number of $u^*$ is at most two (Fig. \ref{fig:number_of_ess}B). If $\gamma>h(u^\dagger)$,  the left-hand side of \eqref{eq:gradient} is negative for all $u^* \in (0,1)$, resulting in the underuse of natural resources ($u\to 0$).
\par
When $w>1$ and $\beta < (w-1)/(2w-1)$, we have $u^\dagger>1$. This implies that $h'(u)>0$ for all $u \in (0,1)$. A unique $u^*$ exists between $0$ and $1$ (Fig. \ref{fig:number_of_ess}C) if and only if
\begin{align}
h(1)=(1-\beta)^w >  \gamma  > 0.
\end{align}
If $\gamma>h(1)$,  the left-hand side of \eqref{eq:gradient} is negative for all $u^* \in (0,1)$, resulting in the underuse of natural resources $u\to0$. In contrast, $\gamma=0$ results in the overuse of natural resources because the left-hand side of \eqref{eq:gradient} is positive for all $u^* \in (0,1)$.
\subsection{Classification of evolutionarily singular strategy}\label{sec:property_ess}
Next, we analyzed the properties of the interior evolutionarily singular strategies. According to inequality \eqref{eq:ES}, an evolutionarily singular strategy is evolutionarily stable if and only if
\begin{align}
    w<1. \label{eq:es_condition}
\end{align}
\par
Inequality \eqref{eq:convergent} shows that an evolutionarily singular strategy is convergence stable when $w<1$. Furthermore, this inequality can be rewritten as follows:
\begin{align} 
&-\alpha w u^* +(w-1)(\mu- \alpha u^*) <0, \nonumber\\
  \Leftrightarrow  &u^*>u^\dagger.
\end{align}
These results clarify that $w$ determines the stability of interior evolutionarily singular strategies.
\par
\begin{figure}[tbh]
    \centering
    \includegraphics[width=0.95\linewidth]{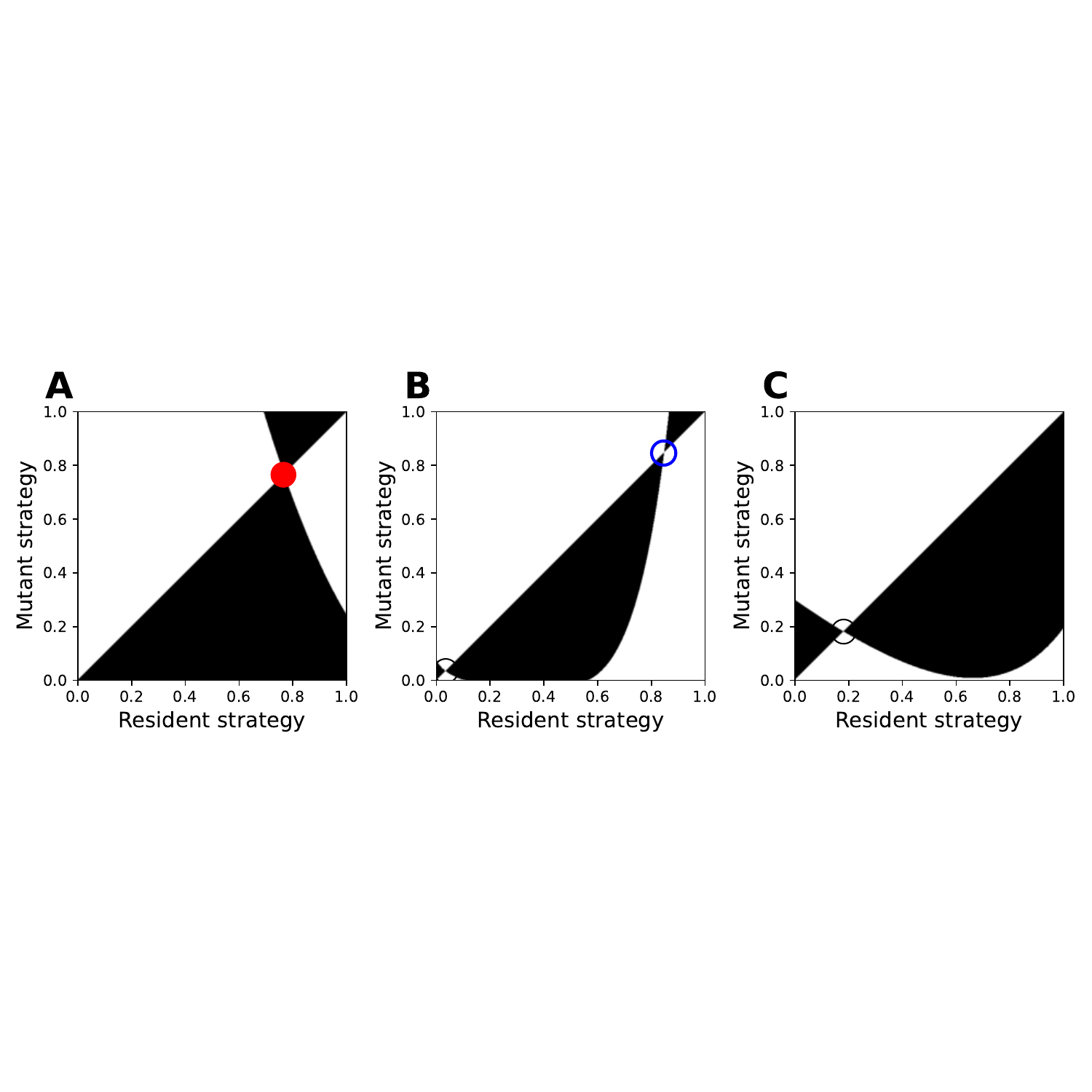}
    \caption{Pairwise invasion plots}
    \label{fig:PIP}
    \begin{flushleft}
    {\small
    Pairwise invasion plots show whether a mutant strategy can invade the population of a given resident strategy (white areas) or not (black areas).  The red-filled circle in panel A shows the CSS, while the blue open circle in panel B represents the branching point. The two open black circles in panels B and C represent the singular points that are neither evolutionarily nor convergent stable.  Parameter values in each panel are as follows: $(b, c, w)=(2, 1, 0.6)$ (A), $(1.667, 1, 1.2)$ (B), and$(0.5, 0.15, 2)$ (C). The other parameter values are fixed as follows: $\mu=1$, $K=200$, $N=100$, and $a=0.005$. 
    
    }
\end{flushleft}
\end{figure}

When $0<w<1$ and $\gamma > h(1)$, the unique interior singular strategy $u^*$ is a continuously stable strategy (Fig. \ref{fig:PIP}A). This evolutionarily singular strategy can also invade nearby strategies when the following inequality is satisfied:
\begin{align}
    \left.\frac{\partial^2 s}{\partial u^2}\right|_{u=v=u^*} > 0
    \Leftrightarrow
    \left\{
    \begin{array}{c}
    \underbrace{\frac{\mu-\alpha u^*}{\mu-3\alpha u^*}}_{>1} >w\\
    \mu>3\alpha u^*
    \end{array}
    \right.
    \quad 
    \mbox{or} \quad 
     \left\{
    \begin{array}{c}
    \underbrace{\frac{\mu-\alpha u^*}{\mu-3\alpha u^*}}_{0<} <w\\
    \mu<3\alpha u^*
    \end{array}
    \right.
\end{align}
\par
When  $w>1$, $\beta > (w-1)/(2w-1)$, and $\gamma < h(u^\dagger)$, two evolutionarily singular strategies exist, but neither of them is evolutionarily stable. Furthermore, only one of them, which is larger than $u^\dagger$ (the blue-open circle in Fig \ref{fig:PIP}B), is convergence stable, and thus it is an evolutionary branching point. Because inequality \eqref{eq:protected} always holds, the two distinct strategies are maintained in the society after the branching. The other evolutionarily singular strategy in this case  (the black open circle in Fig \ref{fig:PIP}B) is not convergence stable, and thus the resident strategy deviates from this singular strategy toward either side (i.e., underuse or the evolutionary branching point).
\par
When $w>1$, $\beta < (w-1)/(2w-1)$, and $0<\gamma<h(1)$, there exists a unique evolutionarily singular strategy, but it is neither evolutionarily nor convergence stable. This is because $w>1$ and $u^*<1<u^\dagger$. As a result, evolutionary dynamics drive the resident strategy to $u=0$ or $ u=1$, depending on the initial value of $u$ (i.e., bi-stability). In other words, either overuse or underuse occurs. 

\subsection{Summary of evolutionary fate}
Our analyses in \ref{sec:number_ess} and \ref{sec:property_ess} clarify that $w$, $\beta$, and $\gamma$ alter the number of interior evolutionarily singular strategies and their stability. Table \ref{tab:evo_fate} summarizes the corresponding evolutionary fates and shows that the nonlinearity parameter $w$ primarily determines the diversity of possible outcomes. When $0<w<1$, adaptive dynamics can lead to either overuse ($u \to 1$) or intermediate use ($0<u^*<1$) of the natural resource, depending on $\gamma$ and $h(1)=(1-\beta)^w$. In contrast, when $w>1$, the evolutionary dynamics exhibit qualitatively richer behavior, including overuse, underuse ($u \to 0$), and evolutionary branching. These results highlight the importance of how benefits from provisioning services increase with resource use.

\begin{sidewaystable}[p]
\centering
\caption{Evolutionary fate and parameter conditions}
\label{tab:evo_fate}
\begin{tabular}{ll}
\hline
Evolutionary fate & Parameter conditions \\
\hline

Overuse ($u \to 1$) 
& $0<w<1$ and $\gamma \le h(1)$ \\

& $w>1$, $\beta<(w-1)/(2w-1)$, and $\gamma=0$ \\[6pt] \hline

Underuse ($u \to 0$) 
&$w>1$, $\beta>(w-1)/(2w-1)$ and $\gamma > h(u^\dagger)$ \\
& $w>1$  $\beta<(w-1)/(2w-1)$ and $\gamma > h( 1)$ \\[6pt] \hline

Bistability & $w>1$, $\beta<(w-1)/(2w-1)$, and $0<\gamma<h(1)$  \\
(overuse or underuse) & \\[6pt] \hline

Intermediate use
& $0<w<1$ and $\gamma > h(1)$ \\[6pt] \hline

Evolutionary branching 
& $w>1$, $\beta > (w-1)/(2w-1)$, and $\gamma < h(u^\dagger)$ $^{a}$  \\[6pt]
\hline
\end{tabular}

\begin{flushleft}
 $^{a}$: In this parameter region, two interior singular strategies exist. The larger one is the evolutionary branching point, while
the smaller one is unstable. If the initial value of $u$ is below the smaller singular strategy, the evolutionary dynamics lead to underuse.
\end{flushleft}
\end{sidewaystable}

\subsection{Deviation from the social optimum}
The purpose of our model is to evaluate the deviation of the evolutionarily singular strategy from the social optimum $u_{\mathrm{opt}}$. Since we assume $0<u_{\mathrm{opt}}<1$, $u\to 1$ and $u\to 0$ correspond to overuse and underuse, respectively.
In the following analysis, we are interested in how large the deviation between the social optimum and the interior evolutionarily singular strategy is. We assume in this subsection that $0<w<1$ and $\gamma >h(1)$ so that the unique evolutionarily singular strategy $u^* \in (0, 1)$ is a CSS (Table \ref{tab:evo_fate}).
In our model, $u^*$ is determined solely by the provisioning service $f(u, R^*(u))$, because the non-provisioning benefits $g(R^*(u))$ depend only on the resident-determined ecological equilibrium and therefore the invasion fitness (see Eq \eqref{eq:invasion}) cancel out the non-provisioning benefits. By contrast, the socially optimal strategy $u_{\mathrm{opt}}$ maximizes the full payoff given by Eq \eqref{eq:target} and thus varies with the weight $d$ of the non-provisioning services. We interpret changes in $d$ as shifts in the societal value of the non-provisioning services. We first analyze limiting cases $d/b \to 0$ and $d/b \to \infty$, and then construct phase diagrams for the sign and magnitude of the gap $u_{\mathrm{opt}}-u^*$ across parameter space.

\subsubsection{Focusing only on the provisioning service}
When $d/b$ is sufficiently small, the payoff function $\phi(u, R^*(u))$ is dominated by the provisioning service :

\begin{align}
    \lim_{d/b\rightarrow 0}\phi (u, R^*(u)) =f(u, R^*(u)).
\end{align}
Below, we write $f(u, R^*(u))=f(u)$ for convenience.
\begin{subequations}
\begin{align}
    f'(u)&= b(aK)^w u^{w - 1} w \left(1- \beta u\right)^{w-1} \left(1-2 \beta u\right) - c,\\
    f''(u)&=b(aK)^w u^{w - 2} w \left(1- \beta u  \right)^{w-2}\left[(4w-2)\beta^2u^2 - (4w-2)\beta u + (w-1)\right]. %
\end{align}
\end{subequations}
Let us denote that $x=\beta u \in (0, 1)$ and 
\begin{align}
    k(x) = (4w-2)x^2 - (4w-2)x + (w-1),
\end{align}
which determines the sign of $f''(u)$. $k(x)$ and thus $f''(u)=0$ has real roots if and only if
\begin{align*}
    &D=(4w-2)^2-4(4w-2)(w-1) > 0\\
    \Leftrightarrow &w > \frac{1}{2}.
\end{align*}
Note that $w=1/2$ has no root.
\par
When $0<w<1/2$, $k(x)<0$ and $f''(u)<0$ for all $x \in (0,1)$. Then, $f(u)$ is concave for all $u \in (0, 1)$. Thus, the unique root of $f'(u)=0$, if it exists between $(0,1)$, maximizes $f(u)$. We denote such an interior root $\tilde{u}_f \in (0, 1)$. Note that
\begin{align}
    \lim_{u \rightarrow 0_+}f'(u)&\rightarrow +\infty,\\
    f'(1)&=b(aK)^w w \left(1- \beta \right)^{w-1} \left(1-2 \beta \right) - c.
\end{align}
Then, $\tilde{u}_f$ exists if and only if
\begin{align}
   & b(aK)^w w \left(1- \beta \right)^{w-1} \left(1-2 \beta \right) - c<0 \nonumber \\
    \Leftrightarrow &
    \gamma> (1-\beta)^{w-1}(1-2\beta). \label{eq:u_f}
\end{align}
If $1>\beta> 1/2$, the above inequality is always satisfied. If $\beta=1/2$, inequality \eqref{eq:u_f} results in $\gamma >0 \Leftrightarrow c>0$. If $0<\beta<1/2$, inequality \eqref{eq:u_f} indicates that the existence of the ess $u^*$  is sufficient for the existence of $\tilde{u}_f$ because $0<1-2\beta<1$; see inequality \eqref{eq:ess_exist1}. 
\par
When $1/2<w<1$, $k(x)=0$ has real roots but they exist outside the interval of $[0, 1] $ because $k(0)=k(1)=w-1<0$. Then, $f(u)$ is concave for all $u\in (0, 1)$, and the analysis of the existence of $\tilde{u}_f$ in the above paragraph holds. 
\par
Now we evaluate the difference between the evolutionarily singular strategy $u^*$ and the social optimum $\tilde{u}_f$. By definition,
\begin{align}
    &f'(\tilde{u}_f)=0 \nonumber \\
    &\Leftrightarrow h(\tilde{u}_f)\frac{1-2\beta \tilde{u}_f}{1-\beta \tilde{u}_f}-\gamma =0
\end{align}
Since $h(\tilde{u}_f)>0$ and $\gamma\geq 0$, it follows that $(1-2\beta \tilde{u}_f)/(1-\beta \tilde{u}_f)\geq 0$, hence 
\begin{align}
\tilde{u}_f\leq \frac{1}{2\beta}.
\end{align}
Note that $0\leq (1-2\beta \tilde{u}_f)/(1-\beta \tilde{u}_f)<1$ because $0<\beta<1$ and $0<\tilde{u}_f<1$. This leads to
\begin{align}
    \gamma = h(u^*)<h(\tilde{u}_f).
\end{align}
Because $h(u^*)$ is a strictly decreasing function when $0<w<1$ (see Eq \eqref{eq:gradient_h}), the above inequality implies
\begin{align}
    u^*> \tilde{u}_f.
\end{align}
In other words, the evolutionary dynamics cause the overuse problem when the non-provisioning benefits are neglected.

\subsubsection{Focusing only on the non-provisioning service}
When $d/b$ is sufficiently large, the payoff is dominated by the non-provisioning benefits:
\begin{align}
    \lim_{d/b\rightarrow \infty}\phi (u, R^*(u)) = g(R^*(u)),
\end{align}
which is maximized at $R^*(u)=R_{\mathrm{opt}}$. Because $R^*(u)$ is a strictly decreasing function of $u$, we denote 
\begin{align}
    R_{\mathrm{opt}}&=R^*(\tilde{u}_g)\nonumber \\
    \Leftrightarrow \tilde{u}_g &= \frac{1}{\beta}\left(1-\frac{R_{\mathrm{opt}}}{K}\right).
\end{align}
Thus, while $\tilde{u}_g$ depends on ecological and demographic parameters through $\beta=\alpha/\mu$ (with $\alpha=aN$) and $K$, it is independent of the parameters affecting the provisioning service $(b,c,w)$ that determine the invasion fitness. These results indicate that all three situations, overuse, underuse, and the optimal use, can happen depending on the value of $R_{\mathrm{opt}}$.

\subsubsection{Balancing the provisioning and non-provisioning benefits}
\begin{figure}
    \centering
    \includegraphics[width=0.95\linewidth]{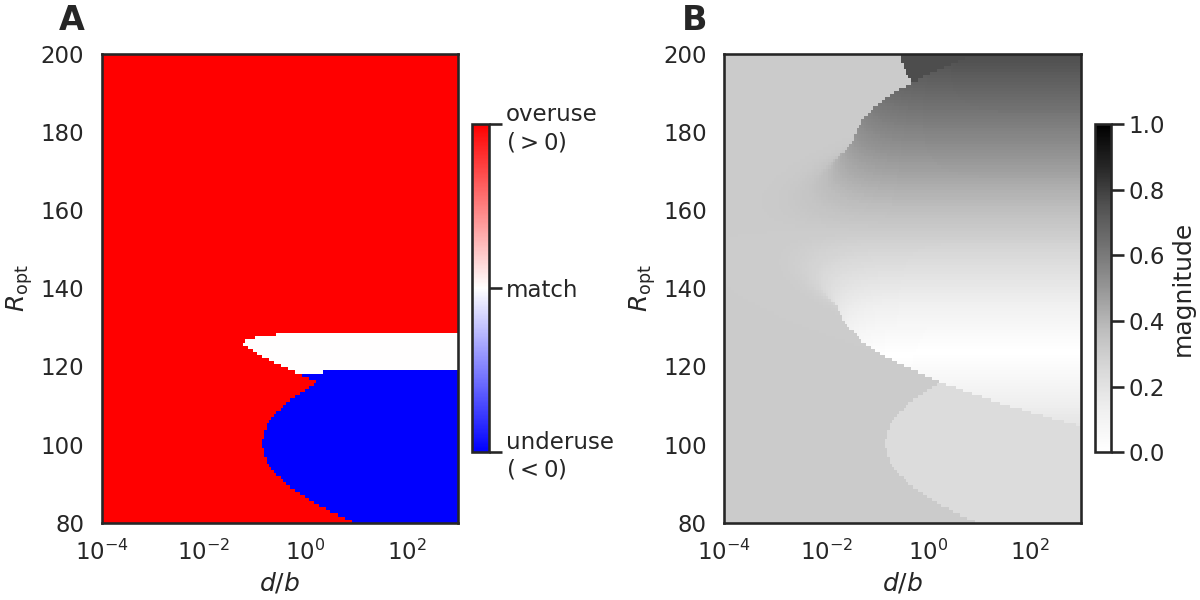}
    \caption{Phase plane analysis of the over- and underuse}
    \begin{flushleft}
        \small{
        The sign of $u^*-u_{\mathrm{opt}}$ (panel A) and the magnitude of the difference (panel B) is shown over $d/b$ and $R_{\mathrm{opt}}$. In panel A, the white area represents that the difference between the evolutionarily singular strategy $u^*$ and the social optimum $u_{\mathrm{opt}}$ is small, $|u^*-u_{\mathrm{opt}}|<0.05$. The red area indicates the overuse, $u^*-u_{\mathrm{opt}}>0.05$ while the blue area shows the underuse $u^*-u_{\mathrm{opt}}<-0.05$. In panel B, the darker area indicates the larger difference between the evolutionarily singular strategy and the social optimum ($|u^*-u_{\mathrm{opt}}|$). Parameter values are as follows: $b=2$, $c=1$, $w=0.6$, $\mu=1$, $K=200$, $N=100$, $a=0.005$, and $e=0.01$.
        
        }
    \end{flushleft}
    \label{fig:phaseplain}
\end{figure}

When the ratio of the non-provisioning to the provisioning benefits is intermediate ($0<d/b<\infty$), the payoff function incorporates both types of benefits. The optimal usage $u_{\mathrm{opt}}$ is then generally difficult to express in a closed form. Instead, we evaluated the deviation $u_{\mathrm{opt}}-u^*$ through the numerical simulations (Fig. \ref{fig:phaseplain}). Consistent with the above two cases, small $d/b$ results in the overuse problem, while large $d/b$ can lead to all three outcomes depending on $R_{\mathrm{opt}}$: overuse (large $R_{\mathrm{opt}}$), match ($u^*\approx u_{\mathrm{opt}}$; intermediate $R_{\mathrm{opt}}$), and underuse (small $R_{\mathrm{opt}}$). These three situations also occurred when $d/b$ was moderate ($ d/b \approx 1$). Changing other parameter values resulted in qualitatively consistent patterns (Figs. S1-S8). 

\section{Discussion}
Most studies of common-pool resources have focused on the problem
of overuse, emphasizing how self-interested exploitation depletes shared
resources \citep{Hardin1968, Ostrom1990Book, Ostrom2008, Frischmann2019}. In contrast, recent empirical studies increasingly report declines or losses of ecosystem services due to insufficient use or abandonment \citep{Shimada2018, Hirahata2020, Brosette2022}. Here, we showed that overuse and underuse are not separate phenomena requiring distinct explanations, but instead emerge naturally as alternative evolutionary outcomes of the same eco-evolutionary process. By applying the adaptive dynamics framework to a simple model coupling resource dynamics and resource-use strategies, we demonstrated that evolutionary dynamics can lead to two qualitatively distinct forms of commons failure: $u\to1$ corresponds to classical overuse, and $u\to0$ represents underuse. Furthermore, our model shows richer evolutionary fates: convergence to intermediate stable strategies or evolutionary branching (Table \ref{tab:evo_fate}). This unified framework provides a theoretical basis for viewing commons dilemmas as intrinsically two-sided problems, in which excessive and insufficient use are both plausible evolutionary fates.
\par
A key result of our analysis is that the qualitative structure of evolutionary outcomes is primarily determined by the nonlinearity parameter $w$, which governs how provisioning services increase with resource use $auR$. For $0<w<1$, evolutionary dynamics are stabilizing and lead either to intermediate use or to overuse, whereas for $w>1$ the system admits a much richer set of
outcomes: overuse, underuse, bi-stability, and evolutionary branching. This highlights the evolutionary mechanism by which the curvature of the provisioning services shapes the selection gradient and alters the evolution of resource use strategies, an aspect overlooked in discussions of underuse. 
\par
Our results also show some consistent patterns with empirical studies. \citet{Shimada2018} point out demographic and socio-economic drivers of underuse. On the one hand, demographic drivers reflect the depopulation in the local areas and aging, which correspond to a decrease in $N$ in our model. When the population size in the society declines, the social exploitation rate of the natural resource $\alpha=aN$  also decreases, resulting in a smaller $\beta$. When $\beta$ cannot exceed the threshold $(w-1)/(2w-1)$, evolutionary dynamics results in either underuse or bistability (Table \ref{tab:evo_fate}).
On the other hand, socio-economic drivers indicate a decline in human reliance on local provisioning services because the benefits of these services decrease and resources are replaced by those imported from outside societies. This corresponds to a decrease in $b$ in our model, leading to larger $\gamma$. When $\gamma$ is larger than $h(u^\dagger)$ or $h(1)$, our analysis indicates the evolutionary underuse  (Table \ref{tab:evo_fate}). The model, therefore, offers a simple evolutionary explanation for underuse patterns commonly attributed to demographic or socio-economic change.
\par
When a unique interior CSS evolves (Fig. \ref{fig:PIP}A), the quantitative measure of overuse and underuse depends on the relative weight of provisioning and non-provisioning services (Fig. \ref{fig:phaseplain}), although the invasion fitness accounts only for provisioning services. Our model shows that the interior CSS results in the overuse when the value on the non-provisioning service is small, while weighting on the non-provisioning service leads the CSS to either overuse, underuse, or appropriate use depending on the optimal resource level for the non-provisioning service $R_{\rm{opt}}$. The distinction between provisioning and non-provisioning services has been emphasized conceptually by \citet{Morino2014}, but has rarely been incorporated into formal evolutionary models. Our model, therefore, associates the
classification of the ecosystem services \citep{MA2005} with the evolutionary dynamics of common-pool resource use.
\par
Although a quantitative evaluation of overuse and underuse under evolutionary branching lies beyond the scope of the present analysis, the phase-plane framework developed here can still be informative under restricted conditions. In particular, if evolutionary branching leads to two approximately symmetric
strategies such that half of the population adopts $u^*+\epsilon$ and the other half adopts $u^*-\epsilon$, the social exploitation rate is $\alpha u^*N$. Under this condition, the ecological equilibrium $R(u^*)$ is well-defined, and the deviation from the social optimum, $u^*-u_{\mathrm{opt}}$, can be evaluated in the same manner as in the case of monomorphic CSS (Fig. \ref{fig:PIP}). More generally, symmetry may break down, or additional strategies may emerge
following evolutionary branching. Analyzing such dynamics requires more sophisticated theoretical tools, such as oligomorphic dynamics \citep{Sasaki-Dieckmann2011,Lion2023} or agent-based
simulations, which represent promising directions for future work.
\par
Because our analysis explicitly evaluates how far evolutionary outcomes deviate from the social optimum, it can provide insight into how socially desirable states (i.e., the white areas in Fig. \ref{fig:phaseplain}A) may become more accessible through the introduction of interventions. For example, institutional governance has been shown to mitigate overuse problems \citep{Ostrom1990Book}. Within the framework of the present model, two types of institutional interventions can be considered.
One class of interventions introduces incentives in the form of taxes, subsidies, or adaptive management schemes. For example, payments for ecosystem services, also called payments for environmental services \citep{Engel2008, Wunder2015}, would move the evolved resource-use strategy $u^*$ toward $u_{\mathrm{opt}}$ when people are taxed under overuse and underuse conditions.
These interventions would change the selection gradient by rewarding or penalizing individual resource-use strategies based on the distance between the current resource level $R$ and the socially optimal level $R_{\mathrm{opt}}$. 
A second class of interventions operates not on behavior directly, but on societal values, for instance, through education, norm transmission, or cultural change that enhances the perceived importance of non‑provisioning services. Within the model, such processes are represented as an increase in the relative weight of the non‑provisioning services, quantified by the ratio $d/b$. This abstraction captures the idea that non‑provisioning benefits, particularly those arising from cultural ecosystem services, are difficult to express in monetary terms \citep{Chan2012BioScience} and are influenced by individual perceptions, experiences, and socio‑cultural contexts, rather than by biophysical conditions alone.
Our results show that shifts in social valuation alone do not necessarily change the qualitative evolutionary fates, but can substantially alter the magnitude and direction of the deviation from the social optimum. This suggests that interventions targeting values, even without modifying individual incentives, may help to adjust CSSs toward socially optimal levels. It would be promising to evaluate how these interventions affect the range of appropriate resource use in future studies.
\par
Of course, our model is based on several simplifications, like other studies. For example, we fixed the population size in the society $N$ over time. In reality, the population size can temporarily change due to migration. To account for such demographic change, future studies can incorporate meta-population dynamics that comprise two or more societies. In such a model, a rational individual would migrate from one society to another based on the expected benefits earned in each society. The demographic change would then alter the maximum social exploitation rate ($\alpha=aN$ in the present model), potentially leading to the underuse problem in a society undergoing depopulation. One way to solve such underuse problems caused by depopulation would be engaging people outside the society in the resource-use activities \citep{Hirahata2020}. On the other hand, people outside the society may have valuations of the provisioning and non-provisioning services that differ from those of local people. Then, the abundance of natural resources and the local valuation may evolve, leading the local eco-evolutionary dynamics to another attractor. Future studies can investigate how such a meta-population framework alters the likelihood of overuse and underuse problems, and which interventions can mitigate them.
\par
In summary, we developed a simple eco-evolutionary model of resource use that reveals
overuse and underuse as alternative evolutionary outcomes of the same
underlying process. By explicitly coupling resource-use strategies with
resource dynamics, we showed rich evolutionary fates: overuse, underuse, bistability, intermediate use, and evolutionary branching. Our analysis identifies the shape of provisioning benefits as a key determinant of these outcomes, highlighting a transition from
stabilizing to polarizing evolutionary dynamics. We further show that the ratio of provisioning and non-provisioning services changes the deviation from the social optimum of resource use to the evolutionary realized one. The results provide a theoretical foundation for viewing commons dilemmas as inherently two-sided problems, in which both excessive and insufficient use can undermine social well-being. This framework offers a baseline for future studies on how interventions may prevent both forms of evolutionary failure and promote socially desirable resource-use strategies.

\section*{Acknowledgements}
This study was supported by Research Institute for Humanity and Nature (RIHN: a constituent member of NIHU) Project No. RIHN14210183 to RN.

\section*{Declaration of generative AI use}
The authors used an AI-assisted language tool (Microsoft Copilot) to improve the clarity and readability of the manuscript, and carefully reviewed all content prior to submission. The authors are fully responsible for the scientific content, interpretations, and conclusions.
\section*{Data availability}
Simulation codes will be publicly available upon acceptance.



\clearpage
\begin{appendices}
\setcounter{page}{1}
\setcounter{table}{0}
\setcounter{figure}{0}
\renewcommand{\theequation}{S\arabic{equation}}
\renewcommand{\thefigure}{S\arabic{figure}}
\renewcommand{\thetable}{S\arabic{table}}
\renewcommand{\thesection}{SI \arabic{section}}
\renewcommand{\bibnumfmt}[1]{[S#1]}
\renewcommand{\citenumfont}[1]{S#1}
\section*{Supporting figures}
\begin{figure}[ht]
    \centering
    \includegraphics[width=0.95\linewidth]{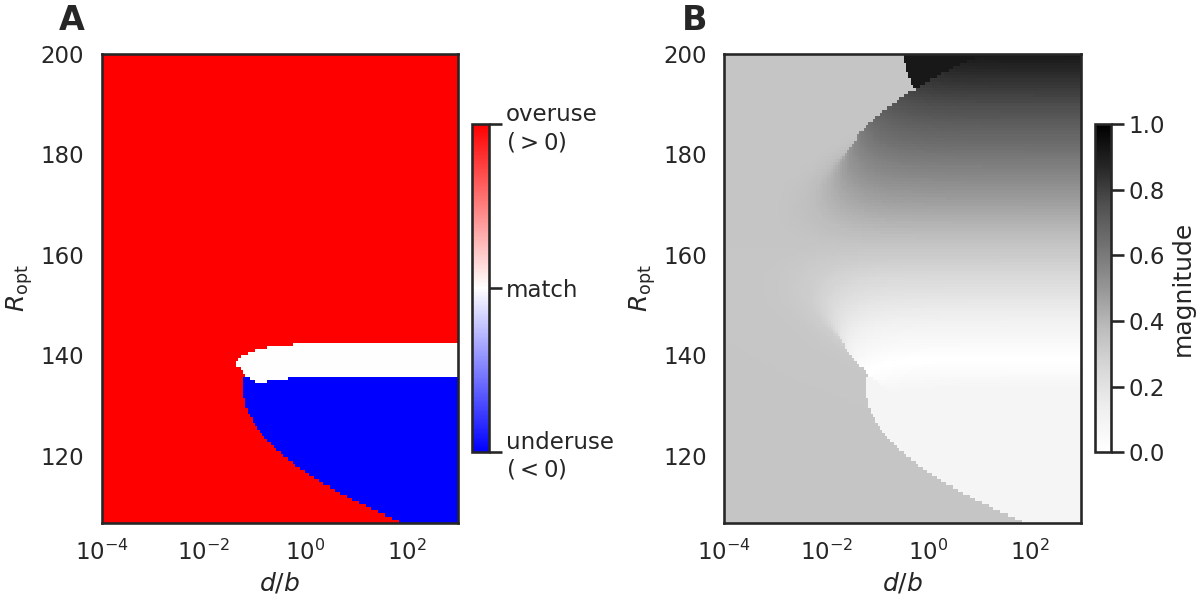}
    \caption{Phase plane analysis with large growth rate}
    \begin{flushleft}
        \small{
        Similar to Fig. 3 in the main text,  the sign of $u^*-u_{\mathrm{opt}}$ (panel A) and the magnitude of the difference (panel B) are shown over $d/b$ and $R_{\mathrm{opt}}$. Here, $\mu=1.5$, and the other parameter values are identical to  Fig. 3.
        }
    \end{flushleft}
\end{figure}

\begin{figure}[ht]
    \centering
    \includegraphics[width=0.95\linewidth]{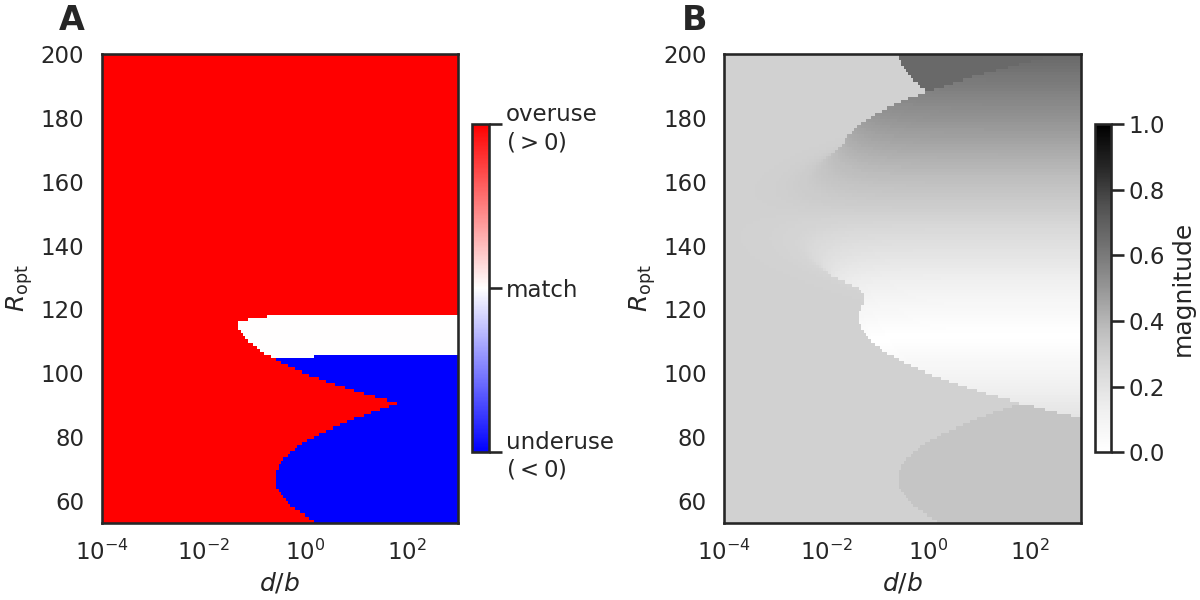}
    \caption{Phase plane analysis with small growth rate}
    \begin{flushleft}
        \small{
        Similar to Fig. 3 in the main text,  the sign of $u^*-u_{\mathrm{opt}}$ (panel A) and the magnitude of the difference (panel B) are shown over $d/b$ and $R_{\mathrm{opt}}$. Here, $\mu=0.75$, and the other parameter values are identical to  Fig. 3.
        }
    \end{flushleft}
\end{figure}

\begin{figure}[ht]
    \centering
    \includegraphics[width=0.95\linewidth]{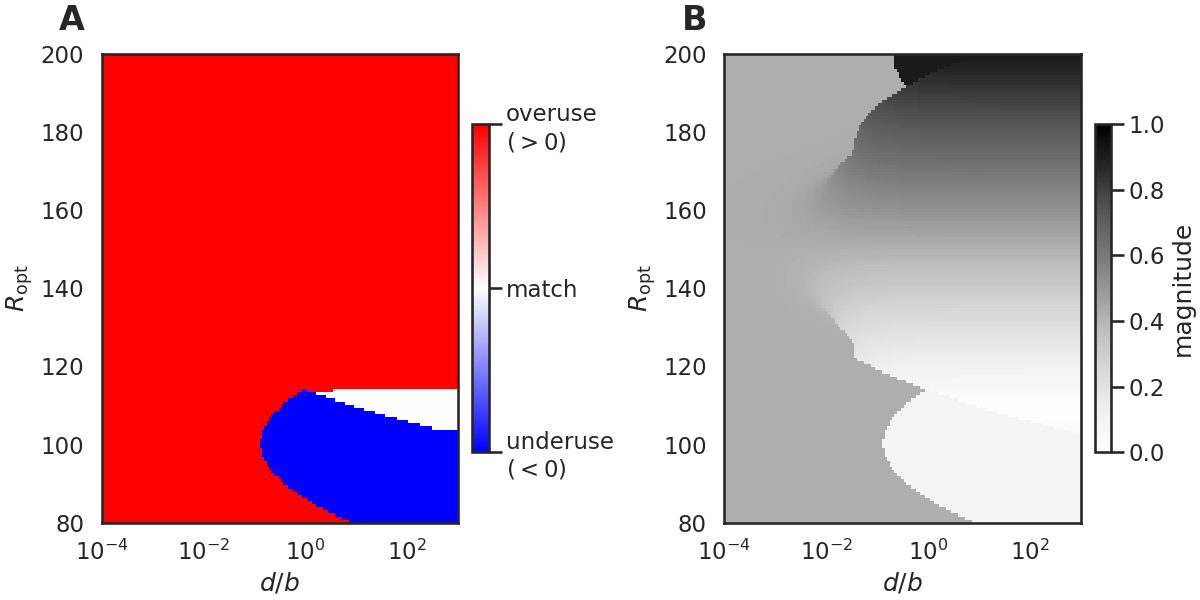}
    \caption{Phase plane analysis with large nonlinearity}
    \begin{flushleft}
        \small{
        Similar to Fig. 3 in the main text,  the sign of $u^*-u_{\mathrm{opt}}$ (panel A) and the magnitude of the difference (panel B) are shown over $d/b$ and $R_{\mathrm{opt}}$. Here, $w=0.8$, and the rest other parameter values are identical to  Fig. 3.
        }
    \end{flushleft}
\end{figure}

\begin{figure}[ht]
    \centering
    \includegraphics[width=0.95\linewidth]{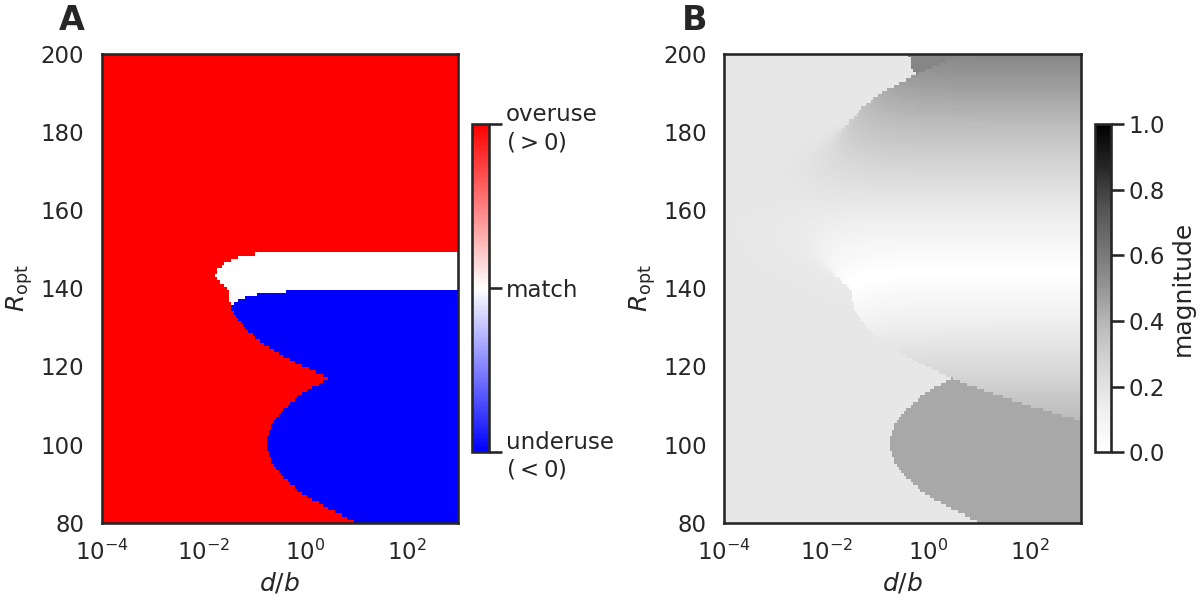}
    \caption{Phase plane analysis with small nonlinearity}
    \begin{flushleft}
        \small{
        Similar to Fig. 3 in the main text,  the sign of $u^*-u_{\mathrm{opt}}$ (panel A) and the magnitude of the difference (panel B) are shown over $d/b$ and $R_{\mathrm{opt}}$. Here, $w=0.5$, and the other parameter values are identical to  Fig. 3.
        }
    \end{flushleft}
\end{figure}

\begin{figure}[ht]
    \centering
    \includegraphics[width=0.95\linewidth]{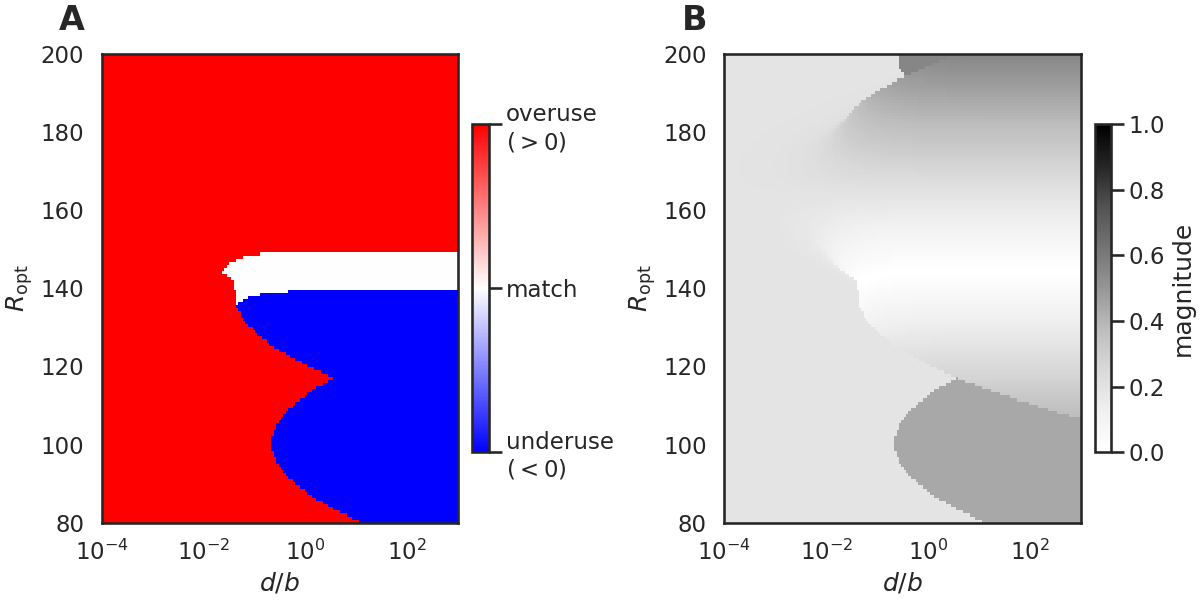}
    \caption{Phase plane analysis with high cost}
    \begin{flushleft}
        \small{
        Similar to Fig. 3 in the main text,  the sign of $u^*-u_{\mathrm{opt}}$ (panel A) and the magnitude of the difference (panel B) are shown over $d/b$ and $R_{\mathrm{opt}}$. Here, $c=1.25$, and the other parameter values are identical to  Fig. 3.
        }
    \end{flushleft}
\end{figure}

\begin{figure}[ht]
    \centering
    \includegraphics[width=0.95\linewidth]{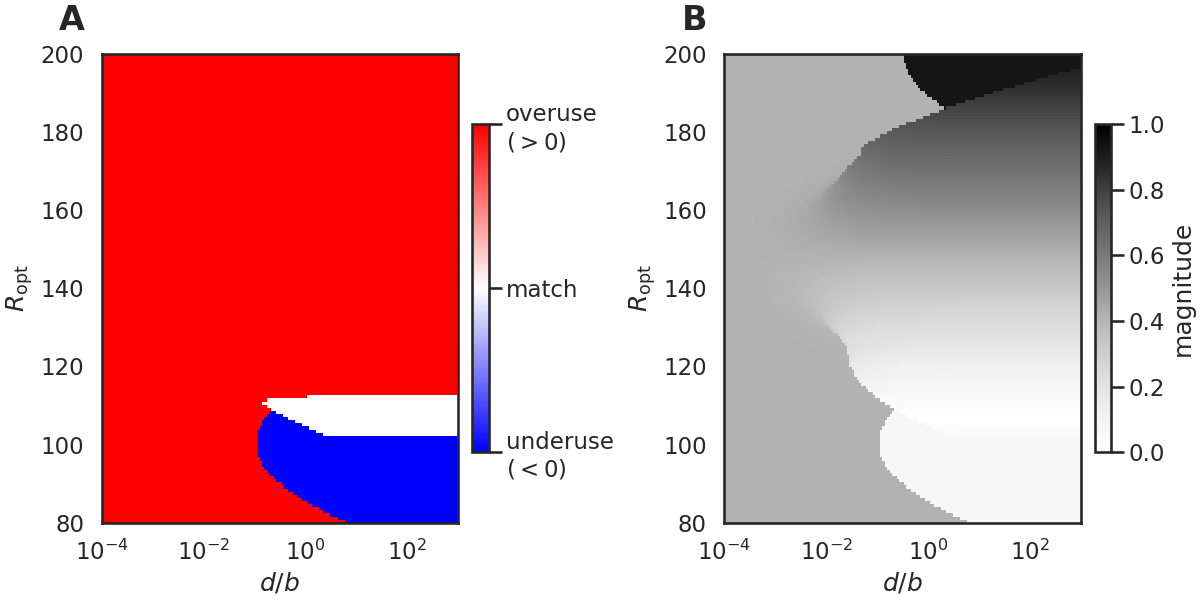}
    \caption{Phase plane analysis with low cost}
    \begin{flushleft}
        \small{
        Similar to Fig. 3 in the main text,  the sign of $u^*-u_{\mathrm{opt}}$ (panel A) and the magnitude of the difference (panel B) are shown over $d/b$ and $R_{\mathrm{opt}}$. Here, $c=0.85$, and the other parameter values are identical to  Fig. 3.
        }
    \end{flushleft}
\end{figure}

\begin{figure}[ht]
    \centering
    \includegraphics[width=0.95\linewidth]{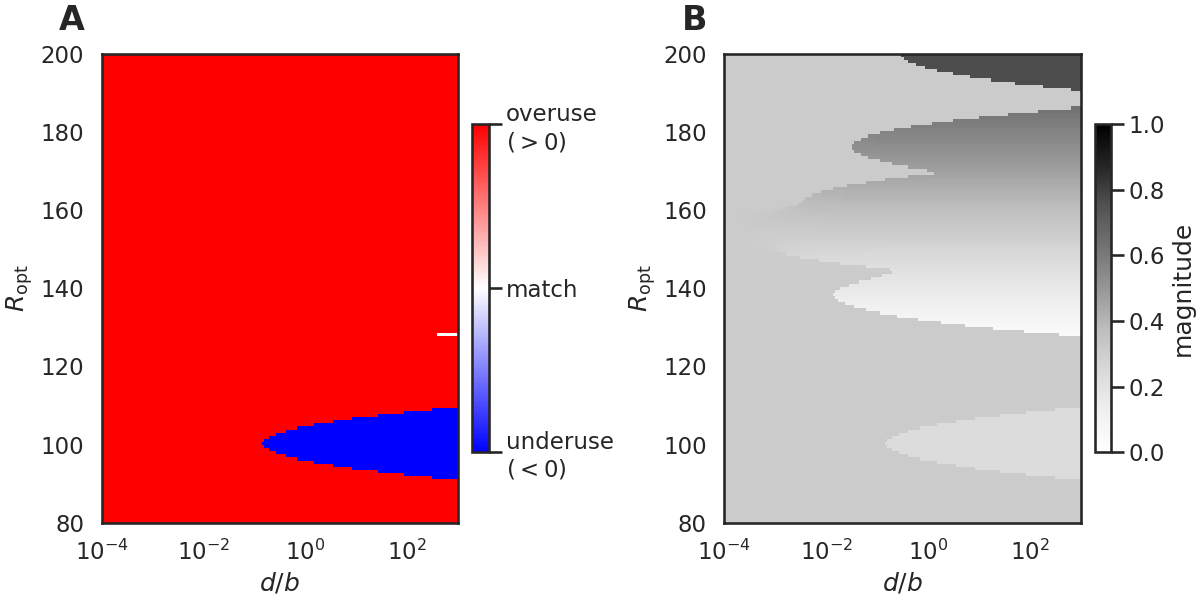}
    \caption{Phase plane analysis with steeper changes in non-provisioning services}
    \begin{flushleft}
        \small{
        Similar to Fig. 3 in the main text,  the sign of $u^*-u_{\mathrm{opt}}$ (panel A) and the magnitude of the difference (panel B) are shown over $d/b$ and $R_{\mathrm{opt}}$. Here, $e=0.1$, and the other parameter values are identical to  Fig. 3.
        }
    \end{flushleft}
\end{figure}

\begin{figure}[ht]
    \centering
    \includegraphics[width=0.95\linewidth]{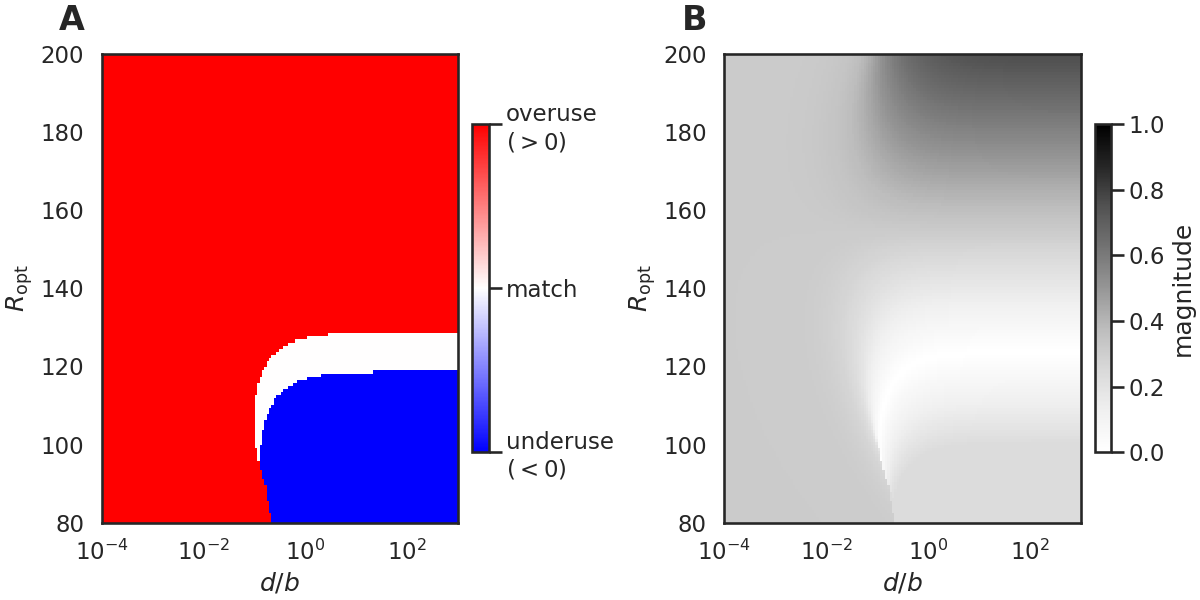}
    \caption{Phase plane analysis with gradual changes in non-provisioning services}
    \begin{flushleft}
        \small{
        Similar to Fig. 3 in the main text,  the sign of $u^*-u_{\mathrm{opt}}$ (panel A) and the magnitude of the difference (panel B) are shown over $d/b$ and $R_{\mathrm{opt}}$. Here, $e=0.001$, and the other parameter values are identical to  Fig. 3.
        }
    \end{flushleft}
\end{figure}
\end{appendices}
\end{document}